\documentclass[final,5p,times,twocolumn]{elsarticle} 

\usepackage{lineno}

\usepackage{graphicx}

\usepackage{amssymb}

\journal{Nuclear Instruments and Methods A}

\begin{document}

\begin{frontmatter}

  \title{The TOP counter and determination of bunch-crossing time at Belle II}

  \author{M. Stari\v c, for the Belle II TOP group}
  \ead{marko.staric@ijs.si}
  \address{J. Stefan Institute, Ljubljana, Slovenia}

  \begin{abstract}
    At the Belle II experiment a Time-of-Propagation (TOP) counter is used for particle identification in the barrel region. This novel type of particle identification device combines the Cherenkov ring imaging technique with the time-of-flight and therefore it relies on a precise knowledge of the time of collision in each triggered event. We discuss the performance of the counter and present a maximum likelihood based method for the determination of event collision time from the measured data.
  \end{abstract}


\end{frontmatter}


\section{Introduction}

The Belle II experiment~\cite{BelleII, TDR} is a second generation of B Factory experiments aimed for the precise measurements in $B$, charm and $\tau$ physics as well as for the searches of physics beyond the standard model. The experiment is sited at KEK, Tsukuba, Japan. The upgraded KEKB collider, the SuperKEKB provides collisions of 4 GeV positrons with 7 GeV electrons at or near the energy of $\Upsilon(4S)$ resonance, which predominantly decays to a pair of $B$ anti-$B$ mesons. With similar cross-sections also pairs of $c \overline{c}$ and $\tau \overline{\tau}$ are produced, enabling to study charm and $\tau$-lepton physics with the same collected data. The SuperKEKB collider utilizes the so called nano-beam optics, with which it is possible to squeeze the beams at the interaction region to a sub-micron dimensions and hence to achieve much larger luminosity at similar beam currents. The SuperKEKB is targeting a 30-times the luminosity of its ancestor.

The Belle II detector is a general purpose spectrometer utilizing charged particle vertexing and tracking, neutral particle detection and particle identification (PID). It consists of the following components: a vertex detector made of two layers of DEPFET sensors (PXD)\footnote{second DEPFET layer is not completely installed yet} and four layers of double-sided silicon detectors (SVD), a central drift chamber (CDC), a time-of-propagation counter (TOP) in the barrel and a proximity focusing aerogel RICH (ARICH) in the forward, both utilizing Cherenkov ring imaging technique, a CsI(Tl) based electromagnetic calorimeter (ECL), and a $K_L$ and muon detector system (KLM). The super-conducting solenoid coil provides a magnetic field of 1.5~T for the charged particle momentum measurements.

Except PXD all other detector components are involved in particle identification: SVD and CDC with energy loss measurements ($dE/dx$), TOP and ARICH exploit Cherenkov ring imaging, ECL is involved with energy deposit measurements and KLM with penetrating power measurements. The last two components mainly contribute to lepton identification, while the first four components contribute mainly to hadron identification. All these components provide log likelihoods for the six stable or long lived charged particles: electron, muon, pion, kaon, proton and deuteron. The log likelihoods are combined by summing over detector components,
\begin{eqnarray}
  \log \mathcal{L}_h = \sum_{\rm det} \log \mathcal{L}_h^{\rm det},~h=\{e, \mu, \pi, K, p, d\}.
\end{eqnarray}
Particle selection is performed by either using a binary PID,
\begin{eqnarray}
  P_{h/h'} = \frac{\mathcal{L}_h}{\mathcal{L}_h+\mathcal{L}_{h'}},
  \label{binaryPID}
\end{eqnarray}
where $h$ and $h'$ denote particles to be distinguished, or by a global PID,
\begin{eqnarray}
  P_{h} = \frac{\mathcal{L}_h}{\sum_{h'}\mathcal{L}_{h'}}.
  \label{globalPID}
\end{eqnarray}
It is also possible to weight the likelihoods in Eq.~\ref{globalPID} with the corresponding prior probabilities.

The Belle II has started taking data in 2019. Since then we recorded 424~fb$^{-1}$, a data sample roughly equivalent to the BaBar or half of the Belle data sample, but compared to the target it represents roughly a 1\% of the final goal. The luminosity has been steadily increasing during past data taking period, reaching a world record of $4.7 \times 10^{34}$~cm$^{-2}$s$^{-1}$ in 2022. To achieve the target of $6 \times 10^{35}$~cm$^{-2}$s$^{-1}$ an increase of the order of magnitude is still needed in the next years.

\section{The TOP counter}

The TOP counter is a variant of the DIRC detector~\cite{DIRC}. Cherenkov photons emitted in a quartz plate by charged particles are transported to the photon detectors by means of total internal reflections. The two dimensional information about the Cherenkov ring is obtained by measuring the time-of-arrival and the position of photons at the photon detectors. The time-of-arrival is measured relative to the $e^+e^-$ collision time and thus includes the time-of-flight of a particle. This kind of DIRC therefore combines time-of-flight measurement with Cherenkov ring imaging technique.

\begin{figure}[h]
  \centerline{\includegraphics[width=7cm]{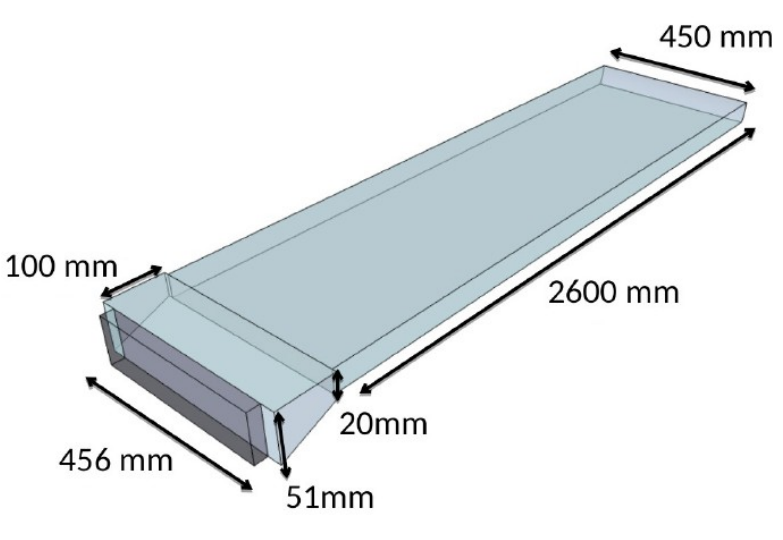}}
  \caption{Quartz optics
    \label{quartz_optics.fig}}
\end{figure}

\begin{figure}[h]
  \centerline{\includegraphics[width=7cm]{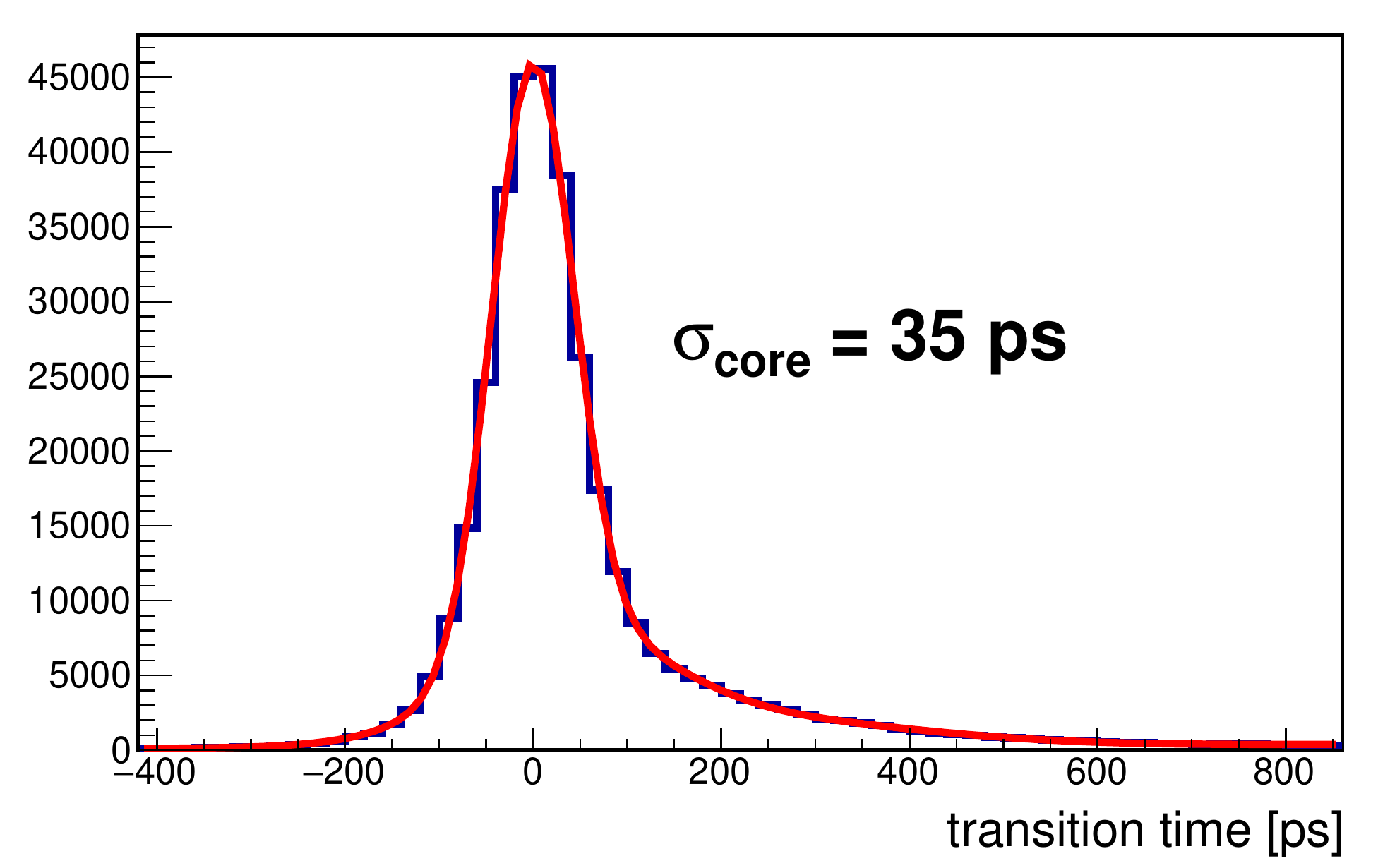}}
  \caption{Transition time spread (TTS) of Hamamtsu R10754.
    \label{tts.fig}}
\end{figure}

The Belle II TOP counter~\cite{Fast:2017} is devoted to hadron ID in the barrel region between polar angles of 32$^0$ and 120$^0$. It consists of sixteen modules positioned at a radius of 120~cm. The quartz optics of a module is composed of a 2.6~m long, 2~cm thick and 45~cm wide quartz plate and a 10~cm long expansion prism at backward side (Fig.~\ref{quartz_optics.fig}). At forward side the quartz plate is shaped to form a spherical mirror of radius-of-curvature of 6.5~m. The prism exit window is equipped with two rows of sixteen Hamamatsu R10754 micro channel plate photo multipliers (MCP-PMT) with NaKSbCs photocathode and $4 \times 4$ anode readout channels~\cite{Hirose:2015} forming an imaging plane of 512 pixels. These tubes are single-photon sensitive, have excellent time resolution( Fig.~\ref{tts.fig}) and can work in a strong magnetic field. 

The readout electronics is based on a 8-channel waveform sampling ASIC developed by the University of Hawaii~\cite{Gary:2019}. Each channel of the chip utilizes a switched-capacitor array with a sampling rate of 2.7 Gs/sec and a 11~$\mu$s long analog ring buffer for storing waveforms. Four ASIC chips are mounted on a carrier board together with a Xilinx Zynq 030-series FPGA which provides clocking and control for the ASICs. A set of four carrier boards and a data aggregator board (SCROD) equipped with a 045-series Xilinx Zynq FPGA form a front-end readout module that interfaces with the Belle II data acquisition system (DAQ). When a trigger is received, the ASIC chips digitize the relevant time interval of the waveforms for triggered channels using 12-bit Wilkinson-type ADC. The digitized data is then sent to the SCROD where the pedestal subtraction and feature extraction (time, amplitude and pulse width) are performed. The feature-extracted data are packed and sent via optical link to the DAQ system. Electronic time resolution of $\sim$50~ps has been obtained for single photon signals.

\section{Calibration of TOP counter}

Calibration of TOP counter involves several steps. At first, the time base of sampling electronics of each of the 8192 electronic channels is calibrated with a precision better than 50~ps (r.m.s). This is performed by injecting double pulses of a constant time delay between the first and second pulse into the inputs. The calibration constants are determined with a minimization procedure described in Ref.~\cite{Staric:2017}. The second step involves time alignment of channels within each module with a precision of at least 50~ps (r.m.s). This is done with a laser calibration system consisting of a pico-second pulsed laser source coupled to a light distribution system made of optical fibers and equipped at output with graded index micro lenses that illuminate MCP-PMT's uniformly as much as possible~\cite{Umberto:2017}. The last two steps are done with muons from $e^+e^- \to \mu^+\mu^-$ events, since particle identities are known in these events. These calibrations involve time alignment of modules and the calibration of bunch crossing time offset~\cite{Staric:2017} with respect to the accelerator RF clock, with which the waveform-sampling electronics is synchronized; the precision is below 10~ps (r.m.s). Besides the timing calibrations we perform also masking of hot and dead channels; the masks are determined from the measured collision data.

The first three calibrations are found to be very stable in time. They are performed at the beginning of each new running period and cross-checked several times during that period. The bunch crossing time offset depends on the accelerator conditions that can change on a daily basis. This calibration is performed continuously for every run.

\section{Particle identification with TOP counter}

Particle identification is based on an extended likelihood method with an analytical construction of the probability density functions (PDF)~\cite{Staric:2008, Staric:2011}. For a given charged particle hypothesis $h$ ($h=e,\mu,\pi,K,p,d$) the extended likelihood is defined as
\begin{eqnarray}
  \log {\cal L}_h = 
  \sum_{i=1}^{N} \log\bigl(\frac{N_h S_h(c_i,t_i)+N_B B(c_i,t_i)}{N_h + N_B}\bigr) \nonumber \\
  + \log P_{N}(N_h + N_B),
  \label{extlkh}
\end{eqnarray}
where $N_h$ and $S_h(c,t)$ are the expected signal yield and signal PDF for the hypothesis $h$, respectively, $N_B$ and $B(c,t)$ are the expected background yield and background PDF, respectively, and $c$ and $t$ are the pixel number and arrival time of the detected photon, respectively. The second term in Eq.~\ref{extlkh} is the Poisson probability to measure $N$ photons while expecting $N_h + N_B$.

The signal PDF for a given pixel $c$ is parameterized as a sum of $m_c$ Gaussian PDF's:
\begin{eqnarray}
  S_h(c,t) = \sum_{k=1}^{m_c} n_{ck} G(t-t_{ck};\sigma_{ck}),
  \label{signal_distr}
\end{eqnarray}
where $t_{ck}$ and $\sigma_{ck}$ are the position and width, respectively, and $n_{ck}$ is the fraction of expected signal photons in the \hbox{$k$-th} peak. Those as well as $m_c$ are determined analytically with the model described in Ref.~\cite{Staric:2011}. The background PDF is modeled as a uniform distribution in a time window in which the photons are measured. The expected background yield $N_B$ is estimated event-by-event from the photon counts of other modules.

\begin{figure}[h]
  \centerline{\includegraphics[width=7cm]{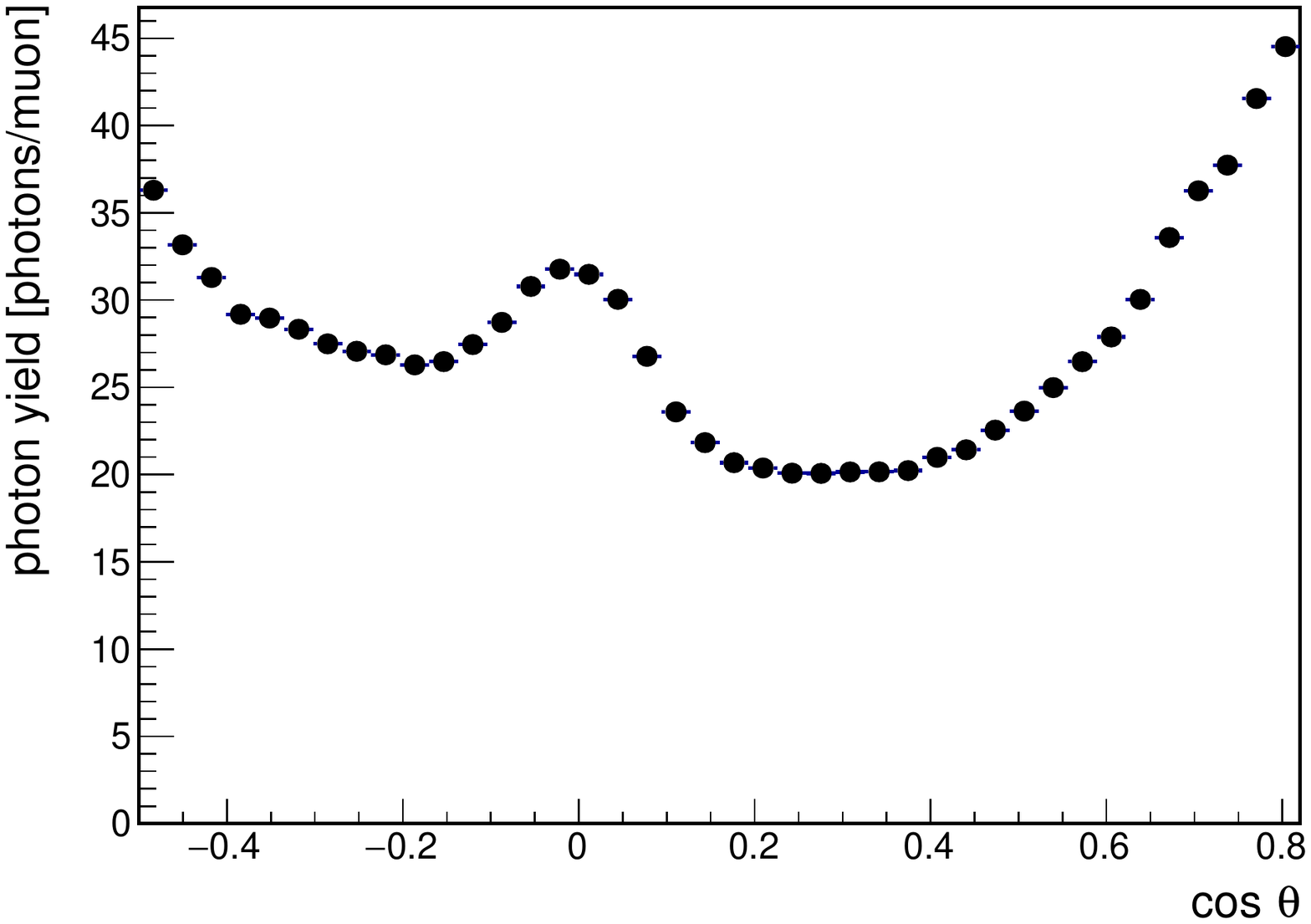}}
  \caption{Number of detected photons per muon as a function of cosine of muon polar angle. 
    \label{nfot.fig}}
\end{figure}

PID performance of TOP counter is governed mainly by two parameters: the number of detected photons per charged particle and the single photon time resolution. Both have been studied with collision data using muons from $e^+e^- \to \mu^+\mu^-$ events. The momentum range of these muons is between 4 and 7 GeV/c, hence the Cherenkov angle is saturated in quartz. The number of detected photons per muon is measured in a time window of 0 to 75~ns, the same as used for the likelihood determination; a time window of -50~ns to 0 is taken to estimate background. Background subtracted photon yields as a function of muon polar angle are shown in Fig.~\ref{nfot.fig}. On average we detect 20 to 45 photons per muon. Strong polar angle dependence is due to several factors: muon trajectory length in the quartz (proportional to $1/\sin \theta$), a fraction of Cherenkov ring satisfying total internal reflection requirement, and the photon losses due to light absorption, quartz surface imperfections and  mirror reflectivity. Photon losses are the largest for polar angles around $\cos \theta \sim 0.3$ since the distance photons must travel is the longest. Enhancement at nearly perpendicular moun impact ($\cos \theta \sim 0$) is due to the fact that the total internal reflection requirement is satisfied for the photons flying directly toward PMT's (direct photons) as well as for those flying toward the spherical mirror (reflected photons).

\begin{figure}[h]
  \centerline{\includegraphics[width=7cm]{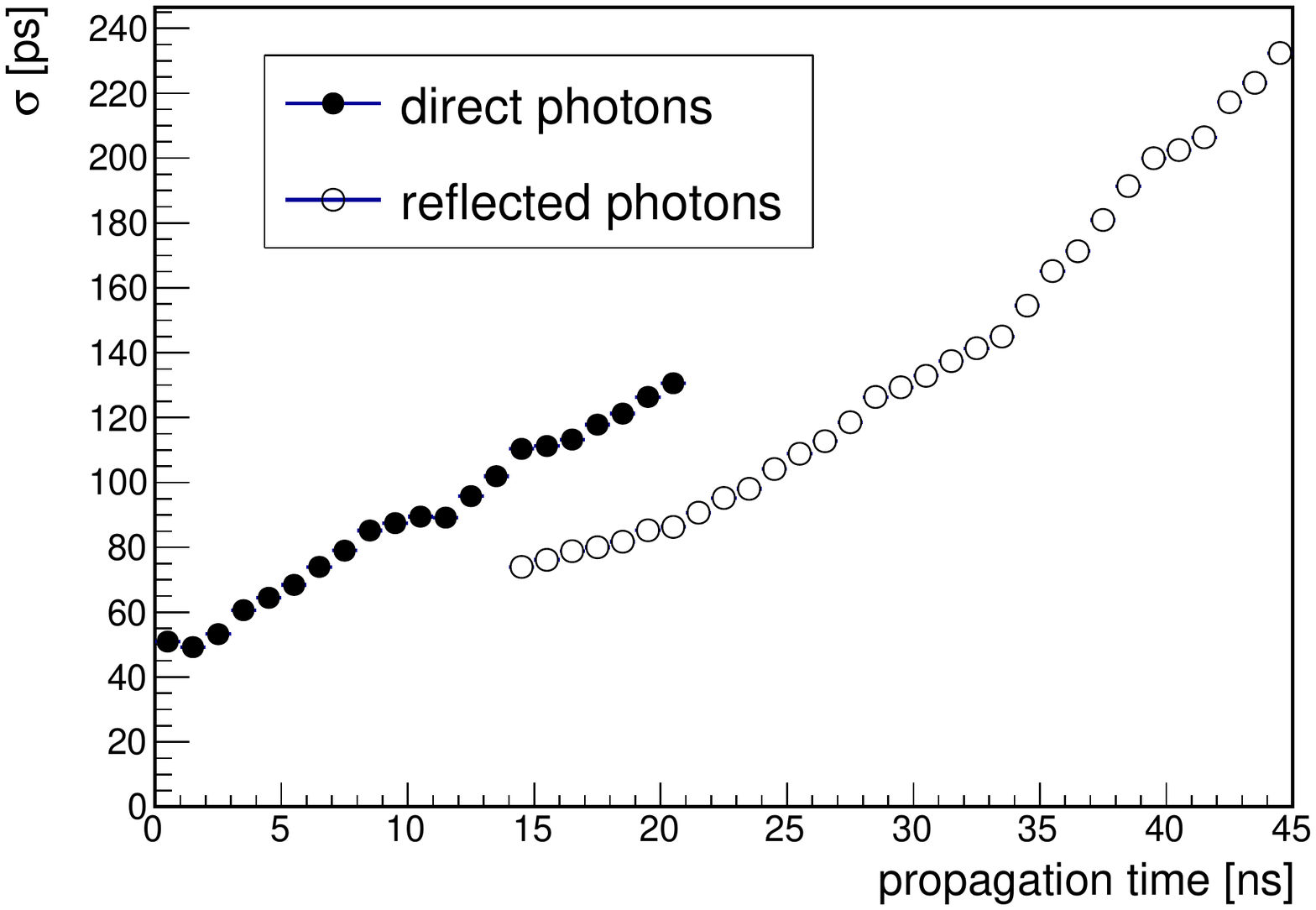}}
  \caption{Single photon time resolution except TTS as a function of photon propagation time.
    \label{time_reso.fig}}
\end{figure}

With muons from $e^+e^- \to \mu^+\mu^-$ events we also measured the time resolution of single photons. The main contribution to the resolution comes from the dispersion of light in quartz (chromatic error) and is proportional to the photon time-of-propagation~\cite{Staric:2008}. We first assigned photons to the peaks of analytic PDF (Eq.~\ref{signal_distr}) using sPlot technique~\cite{sPlot}. The differences of measured photon times and the associated peak positions were then histogrammed in bins of photon propagation time and finally fitted with a convolution of TTS distribution (Fig.~\ref{tts.fig}) and a Gaussian distribution, whose width $\sigma$ is taken as a free parameter. The results are shown in Fig.~\ref{time_reso.fig}. A linear dependence is clearly visible for the direct photons, while for the reflected ones an enhanced time resolution can be noticed; this enhancement is due to chromatic error corrections obtained by focusing photons with a spherical mirror.

\begin{figure}[h]
  \centerline{\includegraphics[width=7cm]{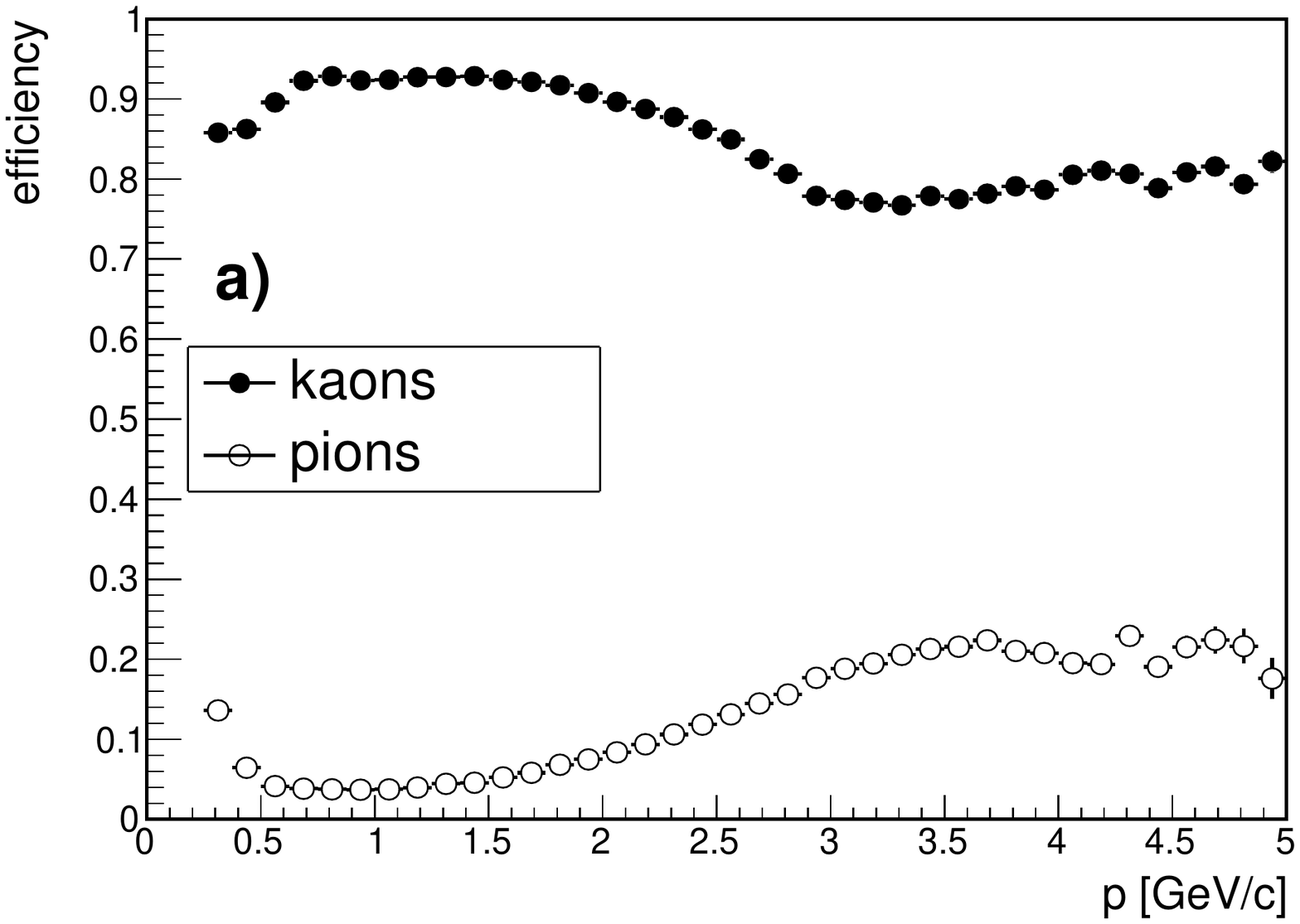}}
  \centerline{\includegraphics[width=7cm]{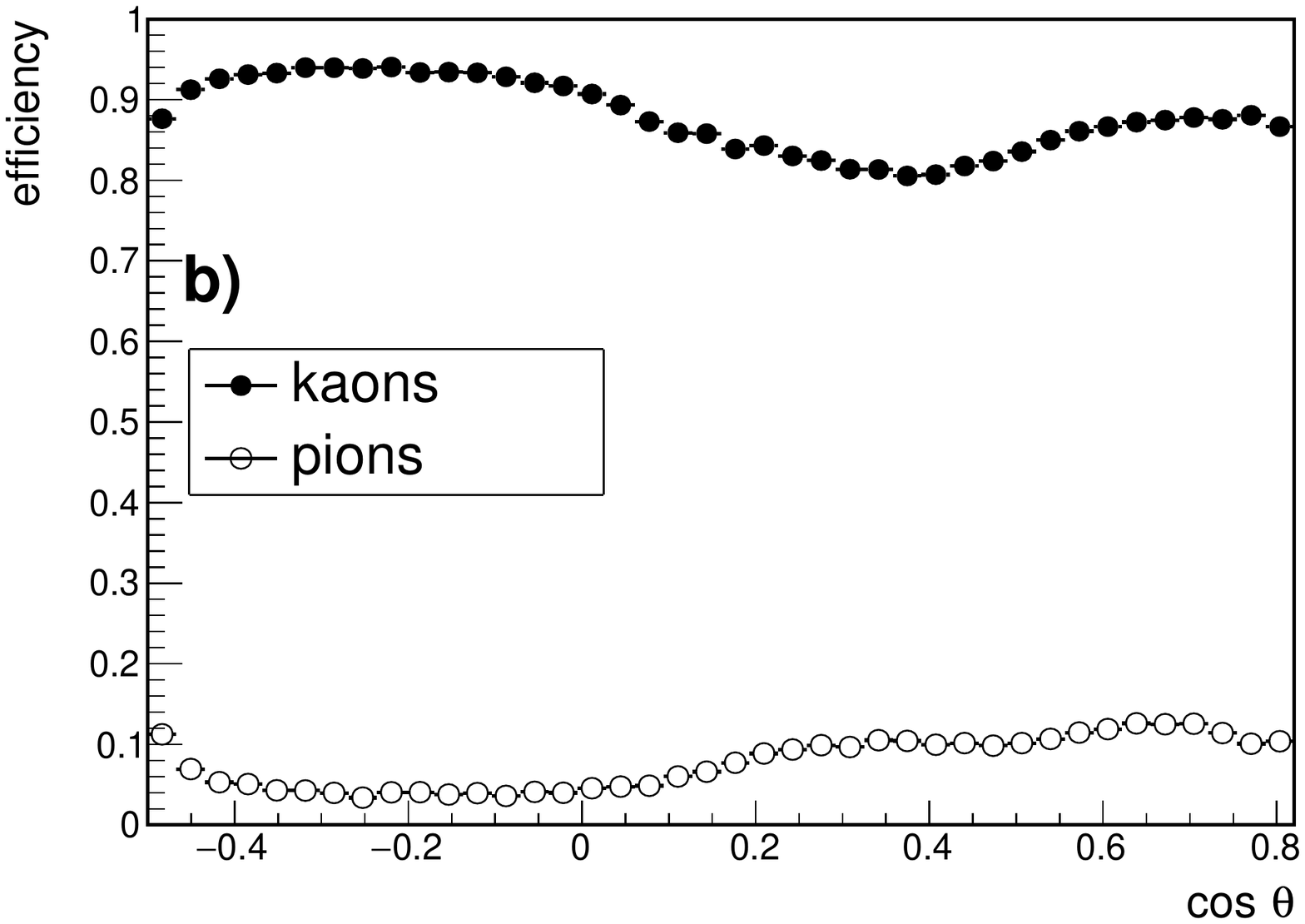}}
  \caption{Kaon efficiency and pion mis-identification probability as a function of momentum (a) and cosine of polar angle (b) for $P_{K/\pi} > 0.5$ as measured with collision data using $D^{*+} \to D^0 (K^-\pi^+) \pi^+$ decays.
    \label{PID.fig}}
\end{figure}

Performance of kaon identification has been studied with collision data using kinematically tagged kaons and pions from $D^0 \to K^-\pi^+$ decays with $D^0$ meson reconstructed in $D^{*+} \to D^0 \pi^+$ decay. The results for $P_{K/\pi} > 0.5$ are shown in Fig.~\ref{PID.fig}. Cherenkov threshold for kaon is at 0.5~GeV/c, while the minimal transverse momentum needed to reach the TOP counter is 0.27~GeV/c. Above the Cherenkov threshold and below 2~GeV/c the identification efficiency is between 90\% and 93\% with 4\% to 8\% pion mis-identification (Fig.~\ref{PID.fig}a). Above 2~GeV/c the performance starts to degrade; at 3~GeV/c it reaches a broad plateau with $\sim$80\% efficiency and $\sim$20\% pion mis-identification. Fig.~\ref{PID.fig}b shows polar angle dependence. In the backward region ($\cos \theta < 0$) the performance is better than in the forward region primarily because of smaller particle momenta. The deep in efficiency at $\cos \theta \sim 0.3$ coincides roughly with the minimum in photon yields shown in Fig.~\ref{nfot.fig}. For these photons also the chromatic error contribution is among the largest.

\section{Determination of bunch-crossing time}

The start for photon time-of-arrival measurements is given by level one trigger whose precision (about 8~ns r.m.s.) does not match the requirement for TOP counter (below 25~ps). This precision can be obtained by identifying a collision bunch-crossing in the off-line processing. The SuperKEKB collider orbits bunches of particles with a frequency of 508~MHz, which corresponds to about 2~ns spacing between RF buckets. The length of a single bunch is 6~mm (r.m.s.), which corresponds to a 14~ps (r.m.s) spread in collision time. If the collision bunch-crossing is uniquely identified, one can correct the measured photon times by the precise timing given with the RF clock and hence can obtain the required start time precision.

The method relies on maximizing the sum of log likelihoods (Eq.~\ref{extlkh}) of particles hitting the TOP counter against a common offset subtracted from the measured photon times. At least one particle in the event that emits enough Cherenkov photons is therefore needed. Particle identities are also required; they are determined from $dE/dx$ measurements in CDC and SVD (the most likely ones are chosen). The result of maximization is then rounded to the nearest RF bucket time and used to correct photon arrival times.

The maximum is searched by scanning a selected time interval because local maxima are usually present. This search is performed in two steps. First, a coarse scan is performed in steps of 0.5~ns within a time interval of $\pm$50~ns using a lookup table of time-projected PDF's. Then a fine scan is performed in a time interval of $\pm$5~ns around the result of the coarse scan, divided into 200 equidistant steps, and using a complete 2-dimensional PDF's. Finally, the maximum is determined precisely by fitting a parabola to the three largest values.

\begin{figure}[h]
  \centerline{\includegraphics[width=7cm]{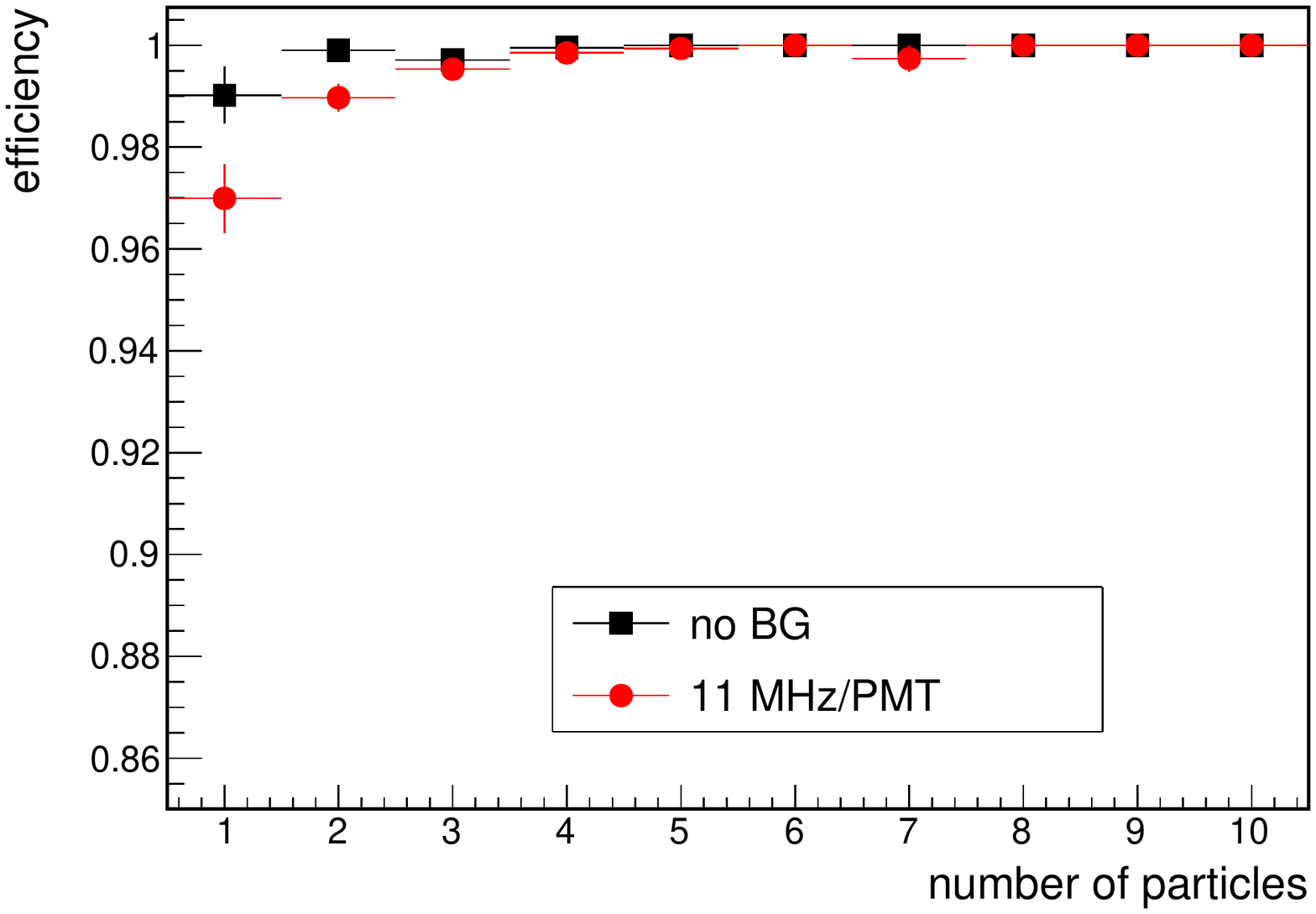}}
  \caption{Efficiency of finding the correct bunch-crossing as a function of particle multiplicity. The average multiplicity of hadronic events is about 4 charged particles in the acceptance of TOP counter.
    \label{bfeffi.fig}}
\end{figure}

Efficiency of finding the correct bunch-crossing depends on particle multiplicity and is found to be very sensitive to beam background. Monte Carlo simulations of generic $B\overline{B}$ events give the following efficiencies: 98.2\% if beam background is absent, 97.4\% with the present background level and 92.1\% with the level expected at the SuperKEKB target luminosity. The inefficiency is found to be primarily due to false maxima caused by Cherenkov photons coming from beam background shower particles. These are not correlated with the collision time, therefore reducing the search interval should increase the efficiency. Recently, SVD can provide the collision time with $\sim$1~ns precision enabling to shorten the search interval. The improved method is using the collision time determined with SVD instead of the coarse scan. In addition, falsely reconstructed bunch-crossings are suppressed by requiring reconstructed bunch-crossing to be matched with a filled bucket. With these modifications the efficiency has been largely improved: 99.9\% at present background level, and 99.5\% at the target luminosity where we expect a background rate of 11~MHz per PMT. The method becomes also much less dependent on particle multiplicity, as shown in Fig.~\ref{bfeffi.fig}.

\section*{Acknowledgments}

We thank the SuperKEKB group for the excellent operation of the accelerator; the KEK cryogenics group for the efficient operation of the solenoid; the KEK computer group for on-site computing support; and the raw-data centers at BNL, DESY, GridKa, IN2P3, and INFN for off-site computing support.

This work was supported by the following funding sources: European Research Council, Horizon 2020 ERC-Advanced Grant No. 884719; Slovenian Research Agency research grants No. J1-9124, J1-4358 and P1-0135.


\bibliographystyle{elsarticle-num}

\end{document}